\font\BBig=cmr10 scaled\magstep2


\def\title{
{\bf\BBig
\centerline{Hopf instantons, Chern-Simons vortices}
\smallskip
\centerline{and}
\smallskip
\centerline{Heisenberg ferromagnets
}
}}

\def\foot#1{
\footnote{($^{\the\foo}$)}{#1}\advance\foo by 1
} 


\def\authors{
\centerline{
P. A. HORV\'ATHY
\footnote{${}^*$}{e-mail: horvathy@univ-tours.fr}}
\smallskip
\centerline{
Laboratoire de Math\'ematiques et de Physique Th\'eorique}
\smallskip
\centerline{Universit\'e de Tours}
\smallskip
\centerline{Parc de Grandmont,
F-37200 TOURS (France)}
}

\def\runningauthors{Horv\'athy}

\def\runningtitle{Hopf instantons, Chern-Simons vortices,
and Heisenberg ferromagnets}


\voffset = 1cm 
\baselineskip = 14pt 

\headline ={
\ifnum\pageno=1\hfill
\else\ifodd\pageno\hfil\tenit\runningtitle\hfil\tenrm\folio
\else\tenrm\folio\hfil\tenit\runningauthors\hfil
\fi
\fi}

\nopagenumbers
\footline={\hfil} 


\def\and{\qquad\hbox{and}\qquad}

\def\smallover#1/#2{\hbox{$\textstyle{#1\over#2}$}} %
\def\2{{\smallover 1/2}}
\def\ccr{\cr\noalign{\medskip}}
\def\parag{\hfil\break} 
\def\={\!=\!}
\def\D{{D\mkern-2mu\llap{{\raise+0.5pt\hbox{\big/}}}\mkern+2mu}\ }
\def\const.{\rm const.}


\newcount\ch 
\newcount\eq 
\newcount\foo 
\newcount\ref 

\def\chapter#1{
\parag\eq = 1\advance\ch by 1{\bf\the\ch.\enskip#1}
}

\def\equation{
\leqno(\the\ch.\the\eq)\global\advance\eq by 1
}

\def\reference{
\parag [\number\ref]\ \advance\ref by 1
}

\ch = 0 
\eq = 1 
\foo = 1 
\ref = 1 


\title
\vskip5mm
\authors
\vskip5mm

\parag
{\bf Abstract.} {\sl
The dimensional reduction of the three-dimensional
model (related to Hopf maps) of
Adam et el. is shown to be equivalent to (i) either the static,
fixed--chirality
sector of the non-relativistic spinor-Chern-Simons
model in $2+1$ dimensions,
(ii)
or a particular  Heisenberg ferromagnet in the plane.
}

\medskip
%

\chapter{Scalar Chern-Simons vortices and Hopf instantons}

In the non-relativistic Chern-Simons  model of
Jackiw and Pi  [1],
one considers a scalar field $\Phi$ which satisfies a second--order
non-linear Schr\"odinger equation,
$$
iD_t\Psi={D_iD^i\over2m}\Psi-g\vert\Psi\vert^2\Psi=0,
\equation
$$
while the dynamics of the gauge field is governed by the Chern-Simons
field/current identities.
 When the coupling constant $g$ is  minus or plus
 the inverse of the Chern-Simons
 coupling constant $\kappa$,
 static solutions arise by solving instead
the self-duality equations,
$$
D_\pm\Psi\equiv (D_1\pm D_2)\Psi=0,
\qquad
(D_k=\partial_k-iA_k),
\equation
$$
supplemented with one of the Chern-Simons equations, namely
$$
\kappa B\equiv\kappa \epsilon^{ij}\partial_iA^j=-\varrho,
\equation
$$
where $\varrho=\Phi^*\Phi$ is the particle density.
 Expressing the gauge potential from
 (1?2) one finds that the other Chern-Simons equations,
 $\kappa E^i\equiv-\kappa\big(\partial_iA^0+\partial_tA^i\big)=
\epsilon^{ij}J^j$,  merely fixes
 $A_t$.
Then, inserting into (1.3) yields the
 Liouville equation, whose well-known solutions provide us with
 Chern-Simons vortices which carry electric and magnetic fields.
The self-dual solutions represent furthermore the absolute
minima of the energy, cf. [1].
\goodbreak

In a recent paper, Adams, Muratori and Nash [2]
consider instead a massless
two-spinor
$\Phi=\pmatrix{\Phi_+\cr\Phi_-}$ on ordinary $3$-space, 
coupled to a (euclidean) Chern-Simons field.
Their field equations read
$$
D_i\sigma_i\Phi=0,
\equation
$$
\null\vskip-15mm
$$
\Phi^\dagger\sigma_i\Phi=B_i.
\equation
$$
Note that this model only contains a (three--dimensional)
magnetic but no electric field.
These authors also mention that
assuming independence of $x_3$ and
 setting $A_3=0$, their model will reduce to the planar self-dual
Jackiw-Pi system, (1.2-3).
 The third component of (1.5) requires in fact
$$
\vert\Phi_+\vert^2-\vert\Phi_-\vert^2=B;
\equation
$$
the two other components imply, however, that either
$\Phi_+$ or $\Phi_-$ has to vanish. Therefore, the reduced equations
read finally one or the other of
$$
D_\pm\Phi_\mp=0,
\qquad
B=\pm\vert\Phi_\pm\vert^2,
\and
\Phi_\mp=0.
\equation
$$
Fixing
up the sign problem by including a Chern-Simons coupling constant
$\kappa$,
these equations look indeed {\it formally} the same as in the self-dual
Jackiw-Pi case. They have, however, a slightly different interpretation:
they are purely magnetic, while those of Jackiw and Pi have a
non-vanishing electric field.
Let us underline that the equations (1.7) differ
from the  second-order field equation (1.1).

\goodbreak
\chapter{Spinor vortices}

Here we point out that the
 model of Adam et al.  reduces
rather more naturally to a particular case of
our spinor model in $2+1$ dimensions  [3].
In this theory, the $4$--component Dirac spinor
with components
$\Phi_-,\,\chi_-,\,\chi_+$ and $\Phi_+$
satisfies the L\'evy-Leblond equations [4]
$$\left\{
\matrix{
(\vec{\sigma}\cdot\vec{D})\,\Phi
&+\hfill&2m\,\chi&=&0,
\ccr
D_t\,\Phi&+\hfill&i(\vec{\sigma}\cdot\vec{D})\,\chi\hfill&=&0,
\cr}\right.
\equation
$$
where  $\Phi$ and $\chi$ are
two-component `Pauli' spinors $\Phi=\pmatrix{\Phi_{-}\cr \Phi_{+}}$
and $\chi=\pmatrix{\chi_{-}\cr \chi_{+}}$. This
non-relativistic Dirac-type equation is
completed with the Chern-Simons equations
$$\eqalign{
&B=(-1/\kappa)\big(\vert\Phi_+\vert^2+\vert\Phi_-\vert^2\big),
\ccr
&E_i=(1/\kappa)\epsilon_{ij}J_j,
\qquad
J_j=i\big(\Phi^\dagger\sigma_j\chi-\chi^\dagger\sigma_j\Phi\big).
}
\equation
$$

In the static and purely magnetic case, $A_{t}=0$,
and chosing $\chi_+=\chi_-=0$,
the second equation in (2.2) is identically satisfied,
leaving us with the coupled system
$$\left\{\eqalign{
&D_+\Phi_-=0,
\cr
&D_-\Phi_+=0,
\cr
&B=(-1/\kappa)\big(\vert\Phi_+\vert^2+\vert\Phi_-\vert^2\big).
\cr}\right.
\equation
$$
Choosing  a fixed chirality,
$
\Phi_-\equiv0
$
or
$
\Phi_+\equiv0,
$
yields furthermore either of the two systems
$$\eqalign{
&D_\pm\Phi_\mp=0,
\cr
&B=(-1/\kappa)\,\vert\Phi_\pm\vert^2,
\cr}
\equation
$$
which, for $\kappa=1$, are precisely (1.7).
For both signs, the equations (2.4) reduce to the Liouville equation;
regular solutions were obtained for
$\Phi_+$ when $\kappa<0$, and for $\Phi_-$ when $\kappa>0$.
They are again purely magnetic,
and carry non-zero spin.

It would be easy keep both terms in (1.7) by allowing a non-vanishing
(but still $x_{3}$-independent) $A_{3}$. Then one would loose the
equations $D_{\pm}\Phi_{\mp}=0$, however.
The impossibility to having both components in (2.3)
but not in (1.7) comes
from the type of reduction performed: while
for spinors one eliminates non-relativistic time, (1.7) comes from
a spacelike reduction. The difference is also related to
the  structure
of the L\'evy-Leblond equation (2.1), which can be obtained by
{\it lightlike} reduction from a
massless Dirac equation in 4-dimensions, while (1.4) comes
by spaceloke reduction  [3].

It is interesting to observe that eliminating
$\chi$ in favor of $\Phi$ in the L\'evy-Leblond equation
(2.3) yields
$$
iD_t\Phi=
\Big[-{1\over2m}\,D_i D^i
+{1\over2m\kappa}\,\big(|\Phi_{+}|^2+|\Phi_{-}|^2\big)\,\sigma_3
\Big]\Phi.
\equation
$$
For both chiralities,
we get hence a second-order equation
of the Jackiw-Pi form (1.1), but with opposite signs
i.e., with attractive/repulsive coupling.

It is worth noting that minima of the energy correspond to
 the coupled equations (2.3) and {\it not} to (2.4).
In fact, the identity
$$
|\vec{D}\Phi|^2=|D_+\Phi_-|^2+|D_-\Phi_+|^2
-(1/2m\kappa)|\Phi|^2\Phi^\dagger\sigma_3\Phi
+\hbox{\rm surface terms}
$$
 shows that the energy of a  field
configuration,
$$
H
=
\int\left\{
(1/2m)|\vec{D}\Phi|^2
+(1/2m\kappa)
|\Phi|^2\Phi^\dagger\sigma_3\Phi
\right\}\,d^2\!\vec{x},
$$
is actually
$$
H={1\over2m}\int d^2\vec{r}
\left\{|D_+\Phi_-|^2+|D_-\Phi_+|^2
\right\},
\equation
$$
which is positive definite, $H\geq0$, provided
the currents vanish at infinity.
The ``Bogomolny'' bound
is furthermore saturated precisely when (2.3) holds.
Its solutions are therefore stable; (2.3) should be considered  as
the true self-duality condition.

\chapter{Heisenberg ferromagnets}

The relative minus sign
of the component densities in the ``provisional'' formula (1.6) differs
from ours in
 (2.3), and is rather that in the $2$-dimensional
Heisenberg  model studied by Martina et al. [5].
Here the spin, represented
by a unit vector ${\bf S}$, satisfies the Landau-Lifschitz equation
$
\partial_t{\bf S}={\bf S}\times\bigtriangleup{\bf S}.
$
In the so-called tangent-space representation, ${\bf S}$ is
replaced by two complex fields, $\Psi_+$ and $\Psi_-$,
 each of which
satisfies a (second-order) non-linear Schr\"odinger equation,
$$
iD_t\Psi_\pm=-\Big[D_iD^i+8\vert\Psi_\pm\vert^2\Big]\Psi_\pm,
\equation
$$
as well as a geometric constraint, $D_+\Psi_-=D_-\Psi_+$.
The covariant derivatives here refer to
 a Chern-Simons-type abelian gauge field,
$$\eqalign{
&B=-8\big(\vert\Psi_+\vert^2-\vert\Psi_-\vert^2\big),
\cr
&E_i=8\epsilon_{ij}J_j,
\qquad
J_i=\big(\Psi_+^*D_i\Psi_+-\Psi_+(D_i\Psi_+)^*)
-\big(\Psi_-^*D_i\Psi_--\Psi_-(D_i\Psi_-)^*\big).
\cr}
\equation
$$
It is now easy to check that in the static and purely magnetic case,
these equations can be solved by the first-order coupled system
$$\eqalign{
&D_\pm\Psi_\mp=0,
\cr
&B=-8\big(\vert\Psi_+\vert^2-\vert\Psi_-\vert^2\big).
\cr}
\equation
$$
For $\Psi_+=0$ or $\Psi_-=0$, we get once again  the equation
of Adams et al..
In the general case, (3.3) leads to an interesting
generalization of the Liouville equation~: making use of its
conformal properties, Martina et al. have shown that it can be
transformed  into the ``sinh-Gordon'' form
$$
\bigtriangleup \sigma= -{\rm sinh} \sigma,
\equation
$$
where $\sigma$ is suitably defined from $\Psi_{+}$ and
$\Psi_{-}$.
Although this equation has no finite-energy regular solution
defined over the whole plane [6], it admits doubly-periodic
solutions i. e. solutions defined in cells with periodic boundary
conditions on the boundary [7].
This generalises the results of Olesen [8]
in the scalar case.
A similar calculation  applied to
the general SD equations, (2.3), of  our spinor model
  would yield
$$
\bigtriangleup \sigma= -{\rm cosh} \sigma,
\equation
$$
whose (doubly periodic) solutions could be interpreted
as
non-linear superpositions of the chiral vortices in [DHP].

\vskip4mm
\centerline{\bf\BBig References}

\reference 
R. Jackiw and S-Y. Pi,
Phys. Rev. Lett. {\bf 64}, 2969 (1990);
Phys. Rev. {\bf D42}, 3500 (1990);
For reviews, see, e. g.,
R. Jackiw and S-Y. Pi, Prog. Theor. Phys. Suppl. {\bf 107},
1 (1992)
or G. Dunne, {\sl Self-Dual Chern-Simons solitons}.
Springer Lecture Notes in Physics (New Series) {\bf 36}, (1995).

\reference 
C. Adam, B. Muratori, and C. Nash,
Phys. Lett. {\bf B 479}, 329 (2000).

\reference 
C. Duval, P. A. Horv\'athy and L. Palla,
Phys. Rev. {\bf D52}, 4700 (1995);
Ann. Phys. (N.Y.) {\bf 249}, 265 (1996).
The same self-dual equations
arise in the relativistic model of
Y. M. Cho, J. W. Kim, and D. H. Park,
Phys. Rev. {\bf D45}, 3802 (1992).

\reference 
J-M. L\'evy-Leblond, Comm. Math. Phys. {\bf 6}, 286 (1967).

\reference 
L. Martina, O. K. Pashaev, G. Soliani,
Phys. Rev. {\bf B48}, 15787 (1993).

\reference
G. Dunne, R. Jackiw, S.-Y. Pi, Trugenberger,
Phys. Rev. {\bf D43}, 1332 (1991).

\reference 
A. C. Ting, H. H. Chen and Y. C. Lee,
Phys. Rev. Lett. {\bf 53}, 1348 (1984);
Physica {\bf 26D}, 37 (1987).

\reference 
P. Olesen, Phys. Lett. {\bf B265}, 361 (1991);
{\sl ibid}. {\bf B268}, 389 (1991).

\end